\documentclass[letter,twoside]{article}
\newcommand{\affil}[1]{$^{\rm #1}$}
\usepackage[authoryear]{natbib}
\bibpunct{(}{)}{;}{a}{}{,}
\usepackage{graphicx}
\date{} %Please leave the date blank
\newbox\grsign \setbox\grsign=\hbox{$>$}
\newdimen\grdimen \grdimen=\ht\grsign
\newbox\laxbox \newbox\gaxbox
\setbox\gaxbox=\hbox{\raise.5ex\hbox{$>$}\llap
        {\lower.5ex\hbox{$\sim$}}}\ht1=\grdimen\dp1=0pt
\setbox\laxbox=\hbox{\raise.5ex\hbox{$<$}\llap
        {\lower.5ex\hbox{$\sim$}}}\ht2=\grdimen\dp2=0pt
\newcommand{\la}{\mathrel{\copy\laxbox}}

\title{\large\bf\flushleft On Cas A, Cassini, Comets and King Charles}
\author{\parbox{\textwidth}{\flushleft
\vspace{-0.5cm}
{\it Roberto Soria\affil{A}, Riccardo Balestrieri\affil{B}, and Yasuyo Ohtsuka\affil{C}}\\
\vspace{0.4cm}
{\small \affil{A}\,International Centre for Radio Astronomy Research, Curtin University, GPO Box U1987, Perth, WA 6845, Australia. Email: roberto.soria@curtin.edu.au}\\
{\small \affil{B}\,Via G. Giacomini 87/14, 47890 Citt\`a, Repubblica San Marino}\\
{\small \affil{C}\,British Library, 96 Euston Road, London NW1 2DB, United Kingdom}}}
\begin{document}
\twocolumn[
\begin{changemargin}{.8cm}{.5cm}
\begin{minipage}{.9\textwidth}
\vspace{-1cm}
\maketitle
%
%
%%%%%%%%%%%%%     ABSTRACT    %%%%%%%%%%%%%
%Abstract of no more than 200 words here.
\small{\bf Abstract:}
We re-examine the long-standing problem of the date of the Cassiopeia A 
supernova (SN), in view of recent claims that it might be the 1630 
``noon-star'' seen at the birth of King Charles II. 
We do not support this identification, based on the expected brightness 
of a Type-IIb SN (too faint to be seen in daylight), the extrapolated 
motion of the ejecta (inconsistent with a date earlier than 1650), 
the lack of any scientific follow-up observations, the lack of any mention 
of it in Asian archives. The origin 
of the 1630 noon-star event (if real) remains a mystery;  
there was a bright comet in 1630 June but no evidence to determine 
whether or not it was visible in daylight. 
Instead, we present French reports about a 4th-magnitude star discovered 
by Cassini in Cassiopeia in or shortly before 1671, which was not seen before 
or since. The brightness is consistent with what we expect for the Cas A SN; 
the date is consistent with the extrapolated motion of the ejecta.
We argue that this source could be the long-sought SN. 

%%%%%%%%%%%%%     KEYWORDS    %%%%%%%%%%%%%
\medskip{\bf Keywords:} History and philosophy of astronomy --- Atlases --- 
Supernovae: general --- Supernovae: individual: Cas A
% Please write all keywords in lower case. PASA uses the
% standard list of subject headings adopted by The Astrophysical Journal
% and available from http://www.journals.uchicago.edu/ApJ/keywords_text.html.
% Keywords are separated by em-dashes, i.e. ---

%%%%%%%%DO NOT EDIT%%%%%%%%%%%%
\medskip
\medskip
\end{minipage}
\end{changemargin}
]
\small
%%%%%%%%EDIT FROM HERE%%%%%%%%%%%%

\section{Introduction}
%Please see the PASA Style Guide for help with correct layout for your manuscript.
%Examples of tables and figures are given below.
The young supernova remnant (SNR) Cassiopeia A is one of the best-studied 
objects in the sky, at all wavelengths; however, the exact date 
of the SN event, as well as the nature of the progenitor, are still 
intriguingly unsolved problems. Finding reliable 17th century evidence 
of visual observations would constrain the peak brightness 
of the event and hence test the currently favoured scenario 
of a relatively faint Type IIb \citep{krause08}, based on the spectral 
study of scattered light echos. 
It would also help modelling the expansion and deceleration of the ejecta, 
and the proper motion of the compact object \citep{fesen06,thorstensen01}.

Historians have long debated the possibility that Cas A was the mysterious 
star ``3 Cas'' observed by John Flamsteed on 1680 August 26 (Gregorian 
calendar), with arguments in favour of this identification 
({\it eg}, \citealt{ashworth80}) because their locations are quite close 
and no other star exists there today, or against it 
({\it eg}, \citealt{stephenson02,green03,stephenson05a}) based on the fact 
that the discrepancy between the position of 3 Cas and Cas A is still 
unusually large for Flamsteed's accurate standards. 
As \citet{stephenson05a} point out, the $10'$ discrepancy between 
3 Cas and Cas A is much larger than Flamsteed's average error of $23''$ for 
the other 20 stars he catalogued on the same night.

In the past couple of years, an alternative suggestion has been presented 
in several international conferences and 
press releases\footnote{See, for example:\\www.ras.org.uk/news-and-press/217-news2011/1948-nam-4-did-a-supernova-mark-the-birth-of-the-merry-monarch \linebreak
and \linebreak chandra.harvard.edu/edu/formal/icecore/king\_charles.html} 
(unfortunately, not in peer-reviewed journals yet) by British astronomer 
Dr Martin Lunn and American historian Dr Lila Rakoczy, 
and has become a topic of discussion also among the educated public 
of amateur astronomers: the possibility that Cas A was the ``noon-star'' 
allegedly visible in the sky on the birth day 
of the future British king Charles II (1630 May 29 in the Julian calendar, 
corresponding to June 8 in the Gregorian calendar).

Moreover, over the last few years, ice core records from Antarctica 
and Greenland have been studied in search of fossil records of impulsive 
ionization events \citep{dreschhoff06,dreschhoff90}. 
As suggested by \citet{rood79}, $\gamma$-rays from SNe can reach 
the polar stratosphere and ionize nitrogen and oxygen, which will then combine 
to form nitrate ions (NO$^{-}_3$). Such nitrates will ultimately be deposited 
in thin layers of polar ice \citep{zeller95}. Two nitrate peaks in the 
Greenland ice core record were interpreted as signatures of Tycho SN 1572 and 
Kepler SN 1604 \citep{dreschhoff06}. In the same ice record, only two 
significant nitrate spikes occur in the second half of the 17th century: 
one around 1667 and one around 1700: \citet{dreschhoff06} suggested 
that either of them could be the signature of Cas A. Other spikes 
were found in the layers corresponding to 1619, 1637, 1639, 1647 
\citep{mccracken01}; no spikes were found for either 1630 or 1680.
However, SN identifications based on nitrate spikes remain a controversial 
topic \citep{motizuki10,motizuki09,risbo81}: gamma-ray bursts, 
soft gamma repeaters, magnetar flares and (most importantly) 
solar proton events can also produce nitrate concentration spikes 
\citep{melott11}, 
and coastal ice records may be contaminated by nitrates transported 
thorough the troposphere from lower latitudes \linebreak \citep{motizuki10}. 
An argument in favour of a non-solar origin (including a possible SN) 
for the 17th century nitrate spikes is the low solar activity 
during the Maunder Minimum ($\approx$ 1645--1715).

%The impulsive event of 1667 during the middle of the Maunder Minimum may represent the very first physical and accurately dated evidence of the supernova remnant Cas A in the terrestrial record.

Given the prominence enjoyed by the noon-star claim on the NASA/{\it Chandra} 
public outreach website and on popular astronomy magazines, 
and the controversial nature of ice-core nitrate events, 
we think it is worth reconsidering this issue. Therefore,  
in this paper, we discuss whether the identification of a 1630 celestial 
event with Cas A is scientifically plausible. We then suggest yet another 
alternative date for the SN, which deserves further investigation.

%http://ssd.jpl.nasa.gov/dat/ELEMENTS.COMET

\begin{figure}
%\begin{center}
\includegraphics[width=7.8cm, angle=0]{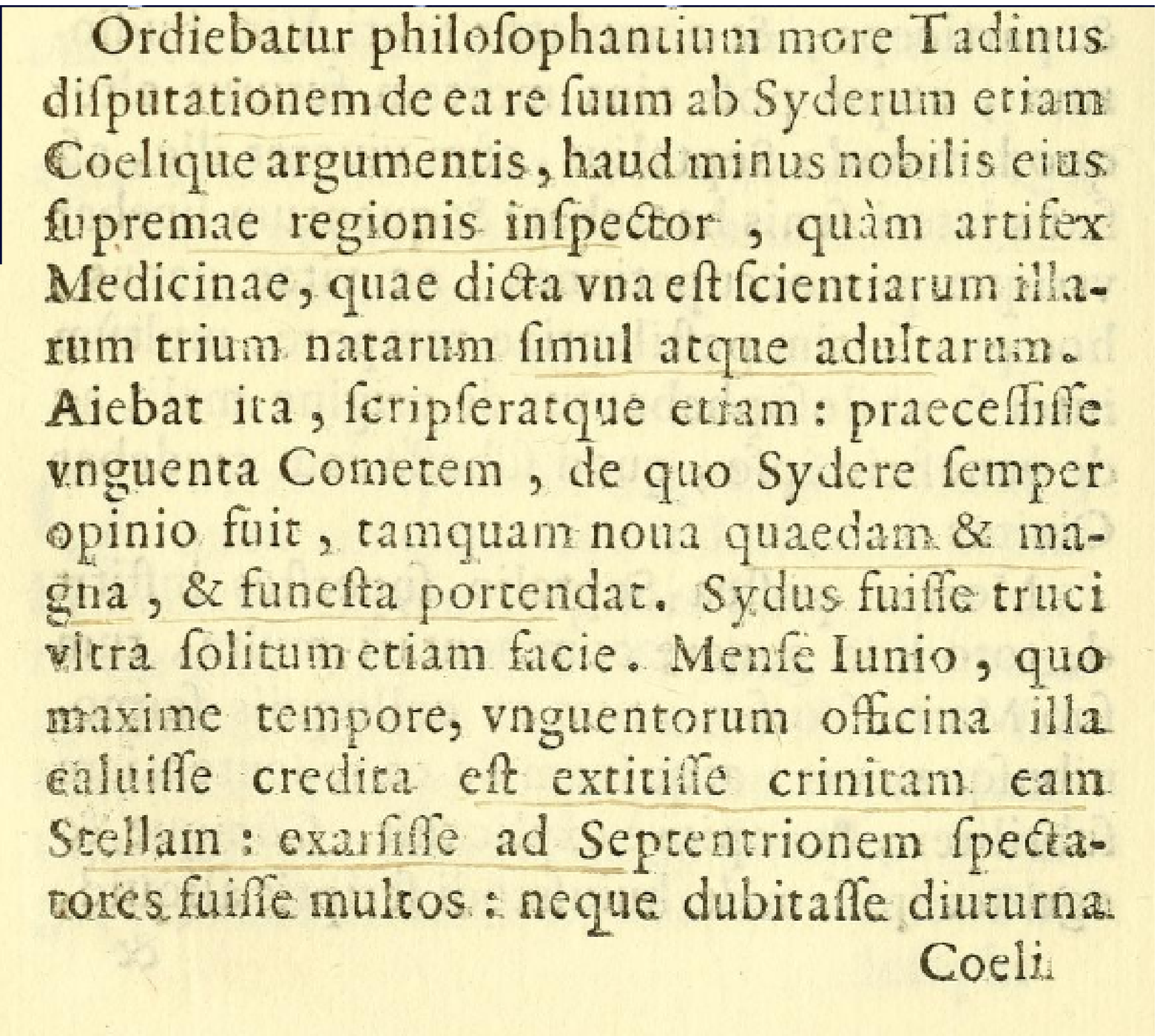}\\
\includegraphics[width=7.8cm, angle=0]{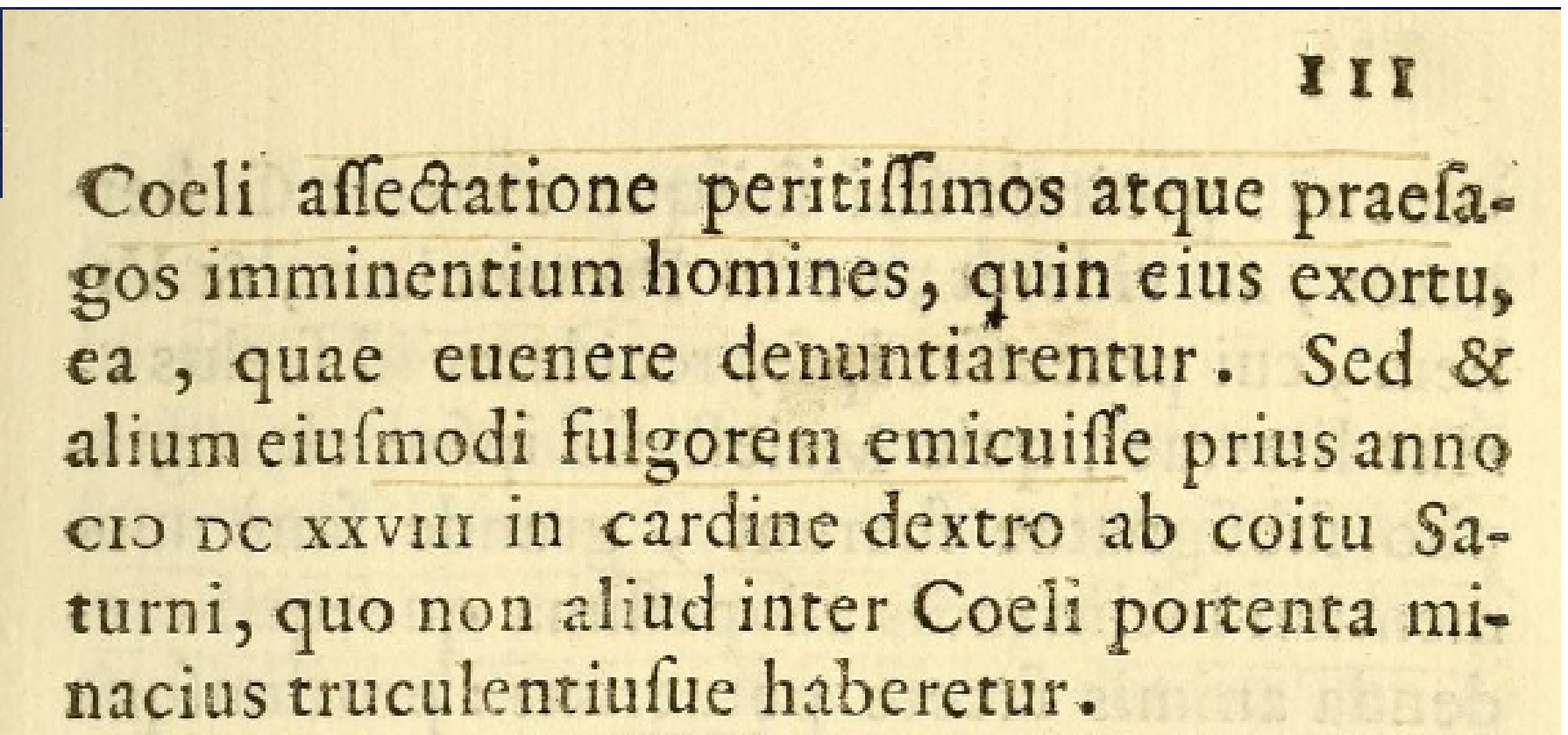}
\caption{Excerpts from Ripamonti's book De Peste, mentioning 
a bright and (apparently) scary comet visible from Milan in 1630 June, 
around the time of King Charles's noon-star. This is the original Latin 
text mentioned in \cite{lynn94}.}
\label{figexample}
%\end{center}
\end{figure}

\section{The 1630 noon-star: comet or supernova?}
Comets are the most obvious candidate for a ``new star''  unexpectedly 
appearing in the sky. Indeed, the 1630 star was sometimes called ``a comet'' 
by contemporary writers and pamphleteers \citep{brown10}.
Lunn \& Rakoczy's argument in favour of 
a supernova rests strongly on their claim that in 1630 
there were no comets bright enough to be visible in 
daylight\footnote{chandra.harvard.edu/edu/formal/icecore/king\_charles.html}. A similar claim was considered and refuted 
by \cite{lynn94}, who cites a rather obscure Latin chronicle 
(Giuseppe Ripamonti's De Peste) describing the scary apparition 
of two great comets in 1628 and 1630, which were believed to be 
responsible for an outbreak of bubonic plague in Milan; 
both comets are also reported by \citet{kronk99}, based 
on the same source.

We examined \cite{ripamonti40}'s original Latin text (Fig.~1): 
%and we agree with \cite{lynn94}'s conclusions. 
Ripamonti does not directly claim to have seen the 1630 comet: he reports 
the observation of the ``physician and natural philosopher''\footnote{This is 
how he described himself on a memorial stone in the now-demolished 
church of Santa Maria della Passarella, in Milan.} 
Alessandro Tadini (1580--1661). According to Ripamonti (p.110), 
Tadini wrote that ``the [comet] star had a scary appearance, 
even more than usual'' (``Sydus fuisse truci ultra solitum etiam facie''), 
a popular description that usually refers to the size 
and shape of a comet's tail. ``This hairy star appeared in the month of June: 
it blazed towards the North, many people saw it'' 
(``Mense Junio [...] extitisse crinitam eam stellam: exarsisse 
ad Septentrionem spectatores fuisse multos''). We checked a later 
edition of Tadini's book \citep{tadini48}, which states: 
``A huge comet appeared near the end of June, towards the North, 
and lasted a long time, 
seen by many people'' (``Apparve nel fine del mese di Giugno una Cometa 
molto grande verso settentrione et dur\`o molto tempo, vista 
da pi\`u persone''), confirming Ripamonti's account, although 
``the end of June'' is up to three weeks later than the London sighting.
%Later on (p.273), Ripamonti mentions again 

Based on those records, we agree with \cite{lynn94}'s conclusions that 
there is good evidence for a bright comet in the summer of 1630, but could 
it be visible in daylight? For that to happen, the head 
of the comet must have an apparent visual magnitude 
$m_v \la -6$ mag, depending on the size of the comet head 
and its angular distance from the Sun. Historically, 
they are rare: we know of less than a dozen comets that reached 
such brightness, and could be seen after sunrise or before sunset. 
Most but not all of them were sungrazing objects 
\citep{sauval97}. For example, Tycho's comet in 1577 
was observed before sunset \citep{kronk99}, even though 
it was not a sungrazing object. Hence, Tadini's observation 
that the comet shone ``towards the North`` is not inconsistent 
with a daylight brightness; alternatively, it may mean that 
its tail was pointing towards the North even though its head 
was near the Sun.
 
In conclusion, a daylight comet in the summer of 1630 cannot be excluded.
The English records are ambiguous on the nature of the object, and there
remains a possibility that the whole story is a legend created to glorify
the Stuart King, as a sign of divine favour\footnote{Similar stories 
are known for other famous rulers. For example, on 1671 August 28, 
``the very night of the marriage [between Czar Peter the Great's parents] 
a brilliant star was perceived quite close to the planet Mars, 
and was thought by the two astrologers to be a good omen'' 
\citep{storksburg88}. In 1941, North Korean Dear Leader Kim Jong-il's birth 
was allegedly heralded by a bright new star in the heavens 
and a double rainbow all the way across the sky 
({\it eg.}, \citealt{french07}).}.
In the Italian record, the memory of the event could be distorted 
by the popular belief that plagues were heralded by comets. 
%(an earlier version of the panspermia theory). 
%The coelestial event in the Chinese records was seen two months later. 
Alternatively, implausible as it may seem, there could have been a bright 
comet and a supernova in the sky at the same time (Cas A would indeed 
be seen towards the North). To make the situation even more confusing, 
there was also an almost total solar 
eclipse\footnote{eclipse.gsfc.nasa.gov/SEsearch/\linebreak 
SEsearchmap.php?Ecl=16300610} 
visible from London an hour 
before sunset on 1630 May 31 Julian Calendar (June 10 in the 
Gregorian Calendar), with a magnitude (that is, the fraction 
of the diameter of the solar disk in eclipse) of 93\%, which 
could have enhanced the perception of astrological significance 
for the alleged noon-star event reported two days earlier.

%\clearpage

\begin{figure}
%\begin{center}
\includegraphics[width=7.8cm, angle=0]{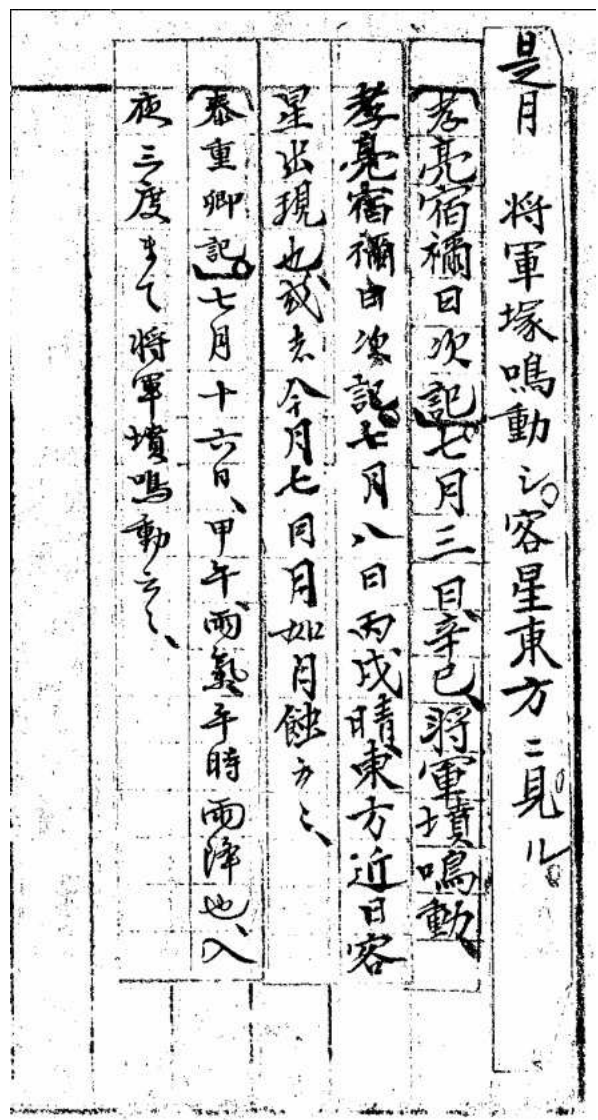}
\caption{Excerpts from the Dai-Nihon Shiryo (Japanese historical annals), 
mentioning the appearance of a guest star on 1630 August 16. 
Image owned by the Historiographical Institute of the
University of Tokyo; courtesy of Prof.~Toru Hoya.}
\label{figexample}
%\end{center}
\end{figure}

\section{Japanese records}
\citet{strom02} cites the apparition of a bright object 
(star or comet) near the Sun, on 1630 August 5, reported by the Chinese 
Ancient Records of Celestial Phenomena \citep{bao88}. 
\citet{strom02}'s favourite interpretation is a sungrazing comet belonging 
to the Kreutz group, seen at perihelion\footnote{Charles II's 
birth star is mistakenly dated 1648 May 29 in \citet{strom02}'s table 1 
and text.}. If the comet perihelion was in August, it is unlikely 
that the same comet could be visible in daylight already in June. 

For an independent test on this and other 17th-century events,
we decided to search for additional information 
in the Japanese archives. We looked into the Dai-Nihon Shiryo Unified 
Database\footnote{www.hi.u-tokyo.ac.jp/english/publication/\linebreak
dainihonshiryo-e.html}, the most complete Japanese historical archives.
%  [大日本史料総合データベース], (URL http://wwwap.hi.u-tokyo.ac.jp/ships/shipscontroller-e) which is maintained by the Historiographical Institute of the University of Tokyo[東京大学史料編纂所]. This is the first place to try for anyone wish to search Japanese historical archives.
Only the chronological history records between AD 887 and 1622 
have been compiled and printed by the Historiographical Institute 
of the University of Tokyo so far; 
however, subsequent records will be published in the next few years. 
%The first volume set of Dai Nihon Shiryo [大日本史料], which covered 887-986, published in 1901. 
We obtained access to yet-unpublished hand-written records 
covering the rest of the 17th century.
%Since then the Historiographical Institute has continued to investigate and research their archive to compile printed chronological history records. Number of volumes which covered up to 1622 has already published by 2012, new volume will be ready for publication in 2013 and so on. 
%The Dai Nihon Shiryo unified Database contains not only published parts, but also unpublished archive which was recorded by hand.
%This hand written copy of archive is categorised highly restricted materials for users, so via the database users can view them in digital format. However this database is very much like a working project, so the database does not give firm grantee to provide every single page of hand written copy of archive in digital format.
We did find a ``guest star'' happening in the year Kan'ei 7th 
(corresponding to 1630). The annals' entry for the 7th month, 
8th day of the Japanese calendar, corresponding 
to August 16, is ``fine weather, a guest star appeared in the East'' (Fig.~2).
%Search result was
%(第一条）山城将軍塚鳴動し、客星、東方に見る
%(Entry 1) Shōgunzuka rumbled, a Guest star appeared in the East.
%*Shogunzuka is a burial mound in Yamashiro which, according to legend, is the burial place of Sakanoue Tamuraamaro. He is regarded as a guardian sprit of the city of Kyoto.  "Rumblings" at this mound were taken as a bad omen. 
%An image of the unpublished hand written copy of archive was attached to the search result and we could find more information to precise the day of events.
%This was associated to the reported rumblings of Shōgunzuka volcano 
We conclude that the Chinese and Japanese records reported the same object, 
but we do not have enough evidence to identify it with the European 
event.

We then searched for other possible events in the Dai-Nihon Shiryo, 
consistent with a SN in Cassiopeia, between 1630 and 1700. We did not 
find any plausible candidates. There are a few objects 
identified as ``broom stars'' (comets), or with corresponding 
detections in China that are identified as comets, 
based on their description of appearance and motion in the sky 
(see also a discussion on the meaning of broom stars in 
\citealt{stephenson05a,stephenson05b}). 
For example, one seen in China on 1668 March 3--12, and reported 
in the Dai-Nihon Shiryo entry of 1668 March 8; another one reported 
in the Chinese records on 1679 September 2 and in the Japanese archives on 
1679 August 21.

An interesting pair of (likely) separate events is mentioned 
in the entry for the year Sh\={o}h\={o} 4th,  
5th month, 25th day, which translates as 1647 June 27 in the Gregorian 
calendar (a year corresponding to a Greenland nitrate spike, 
\citealt{mccracken01}): 
``a mysterious bright object flew in the North-Eastern sky; also, 
a guest star appeared in the West, something I have never seen 
in my life before''. However, there is no evidence to connect either 
of these two events (the first of which is perhaps a bright meteor) with Cas A.
A bright new object reported to have been seen in the East-North East 
sky at dawn on 1661 July 17 could be another sun-grazing comet.
Similarly, on 1671 November 29, a ``hakki'' (a word usually referring 
to meteors or comets) appeared ``between 7 and 9pm in the Western sky, 
looking like a bright column of light'': but that is obviously not 
where Cas A would have been located.
As for the year 1680 (alleged Flamsteed detection), the only  
celestial event in the Japanese annals is, as expected, the Great Comet 
observed in December, well reported in European and North American 
historical records.
%\\  
%{\bf Hatsu please do one final check that we have not missed 
%any interesting guest star specifically between 1667 and 1671}.

%\section{Additional evidence against the 1630 identification}
\section{Brightness of Type IIb SNe}
A key assumption of the 1630 noon-star identification 
is that the Cas A SN must have been bright enough to be seen 
in daylight, as was the Tycho SN in 1572. We now discuss 
whether this is the case.
Optical spectroscopic studies of the light echo from Cas A have revelead 
\citep{krause08} that the supernova was of Type IIb. 
This means that it started as a Type II (with hydrogen lines) but quickly 
evolved to a Type Ib (no hydrogen lines). This type of events 
was recognized as a distinct class only a few years ago, and its physical 
interpretation is still disputed \citep{claeys11}. 
It appears that the progenitor star has only a thin layer of hydrogen left, 
with a mass $M_{\rm H} \sim 0.1$--$0.5 M_{\odot}$ when it collapses.
In one scenario \citep{podsiadlowski93}, the progenitor star 
fills its Roche lobe in a binary system and transfers most 
of its hydrogen envelope to its massive companion star. 
Another scenario \citep{nomoto93} suggests that two massive stars 
in a binary system actually merge, forming a common envelope, which 
is mostly but not entirely removed by the energy released by 
the in-spiralling of the two cores, before the implosion of the merged core.
Observationally, a few percent of core-collapse SNe are now identified 
as Type IIb, with wildly discrepant estimates going from 1.5\% 
to more than 10\% ({\it eg}, \citealt{arcavi10,smartt09,smith11,claeys11} 
for a review). 

There are 76 SNe identified as Type IIb in the Asiago 
Catalogue\footnote{Most updated version available online:\\
heasarc.gsfc.nasa.gov/W3Browse/star-catalog/asiagosn.html} 
\citep{barbon99}; we also cross-checked the Asiago list 
with the List of Supernovae web catalogue hosted by the Harvard-Smithsonian 
Center for 
Astrophysics\footnote{cfa-www.harvard.edu/iau/lists/Supernovae.html}. 
Sixty-eight of them have a host galaxy identification and a discovery 
or peak visual brightness. We converted the apparent brightness 
to absolute magnitudes, using the cosmology-corrected luminosity distances 
in NED\footnote{ned.ipac.caltech.edu}, which are based on the 3K cosmic 
microwave background frame. 
We also corrected each source for line-of-sight extinction, using the values 
from \citet{schlafly11}. When the optical band is not well defined 
or specified in the Asiago Catalogue (as is the case for most SNe discovered 
by amateur astronomers), we used the extinction in the V band.
For a few well-studied SNe 
(1993J: \citealt{schmidt93}; 1996cb: \citealt{qiu99}; 
2001ig: \citealt{bembrick02}; 2008ax: \citealt{pastorello08}; 
2011dh: \citealt{arcavi11}), we used more accurate peak brightnesses 
from individual studies in the literature, rather than the values listed 
in the Asiago Catalogue, but the difference is generally small and 
does not affect our general conclusions. 
We plot the resulting absolute brightness distribution in Fig.~3. 
The distribution is clearly peaked at $-17.5 \la M \la -16.5$ mag. 
The tail of the distribution at fainter magnitudes is probably due 
to objects being discovered past their peak brightness. The small number 
of sources with $-18.5 \la M \la -19$ mag may belong to a different 
sub-class of Type IIb SNe, or may be mis-classifications, 
but that is beyond the scope of this paper. We are also aware that 
the catalogue contains a mixture of discovery and peak brightnesses, 
and often non-standard photometric bands or visual estimates; however, 
the main point of our exercise is simply to show that most Type IIb SNe 
reach a characteristic absolute brightness $\sim -17$ mag. 
(See \citealt{richardson06,richardson02} for the brightness distribution 
of other classes of SNe, based on the Asiago Catalogue.) 
The distance of Cas A is $3.4^{+0.3}_{-0.1}$ kpc \citep{reed95}, 
that is a distance modulus $= 12.7^{+0.2}_{-0.1}$ mag. Adding an extinction 
$A_V \sim 8$ mag \citep{krause08} results in an apparent peak brightness 
$m_V \sim 3$--$4$ mag for Cas A, clearly not visible in day light, 
and not particularly impressive at night, either.

%The first maximum of the light curve is due to shock heating of the thin envelope, while the second maximum is due to the radioactive decay of Ni-56.

% It is preferable to embed your figures in the text as in the following example
\begin{figure}
%\begin{center}
\includegraphics[width=5.8cm, angle=270]{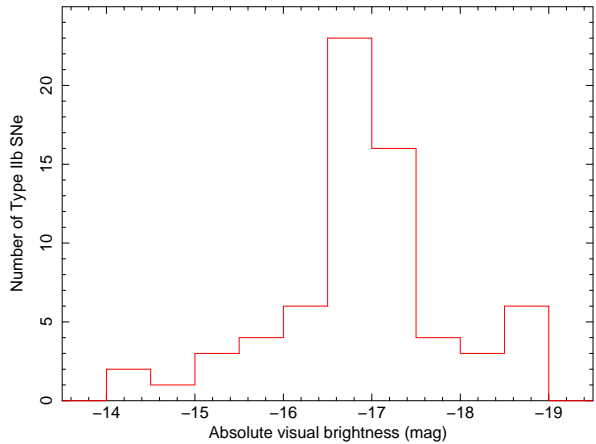}
\caption{Absolute visual magnitude of the 68 Type-IIb SNe identified 
since 1993 with a reliable host galaxy identification. Peak brightness 
were used whenever possible; otherwise, we adopted discovery brightnesses 
from the Asiago catalogue.}\label{figexample}
%\end{center}
\end{figure}

\section{Additional evidence against a 1630 Cas A identification}
Another constraint on the explosion date comes from analyses of the proper 
motion of the ejecta. Such studies have consistently indicated a date 
later than 1650. More specifically, \citet{thorstensen01}
extrapolated an undecelerated convergence date of $1671.3 \pm 0.9$, 
from a sample of 17 bright knots for which archival imaging data 
was available over at least 50 years.
Using a larger sample of 72 bright and/or compact knots imaged 
by the {\it Hubble Space Telescope},
\citet{fesen06} estimated an explosion date of $1671.8 \pm 17.9$ 
(again, neglecting deceleration). Using instead a subsample of knots 
from the northwestern limb, which appear to have suffered the least amount 
of deceleration, \citet{fesen06} estimated $1680.5 \pm 18.7$.

Finally, we need to consider a common-sense argument. If the 1630 day-time 
object was real, and seen by so many common people, it must have caught 
the attention also of natural philosophers, astrologers, astronomers, 
mathematicians, theologians. When a "new star" (Tycho's SN) appeared 
in Cassiopeia in 1572 \citep{brahe73}, 
it sparked tremendous interest, with philosophical
and scientific discussions about the nature of fixed stars, for the first time 
unequivocally showing signs of change rather than being perfect and eternal. 
The new object was immediately reported in sky charts and
mentioned in letters and academic treaties. A similar "new star" event 
in 1630 (in the same region of sky!) would have been equally well reported, 
or more, considering the progress of astronomy in the intervening
six decades. That was clearly not the case; the lack of interest 
is consistent with the 1630 event being a more familiar 
event (meteor or comet).

\begin{figure}
%\begin{center}
\includegraphics[width=7.8cm, angle=0]{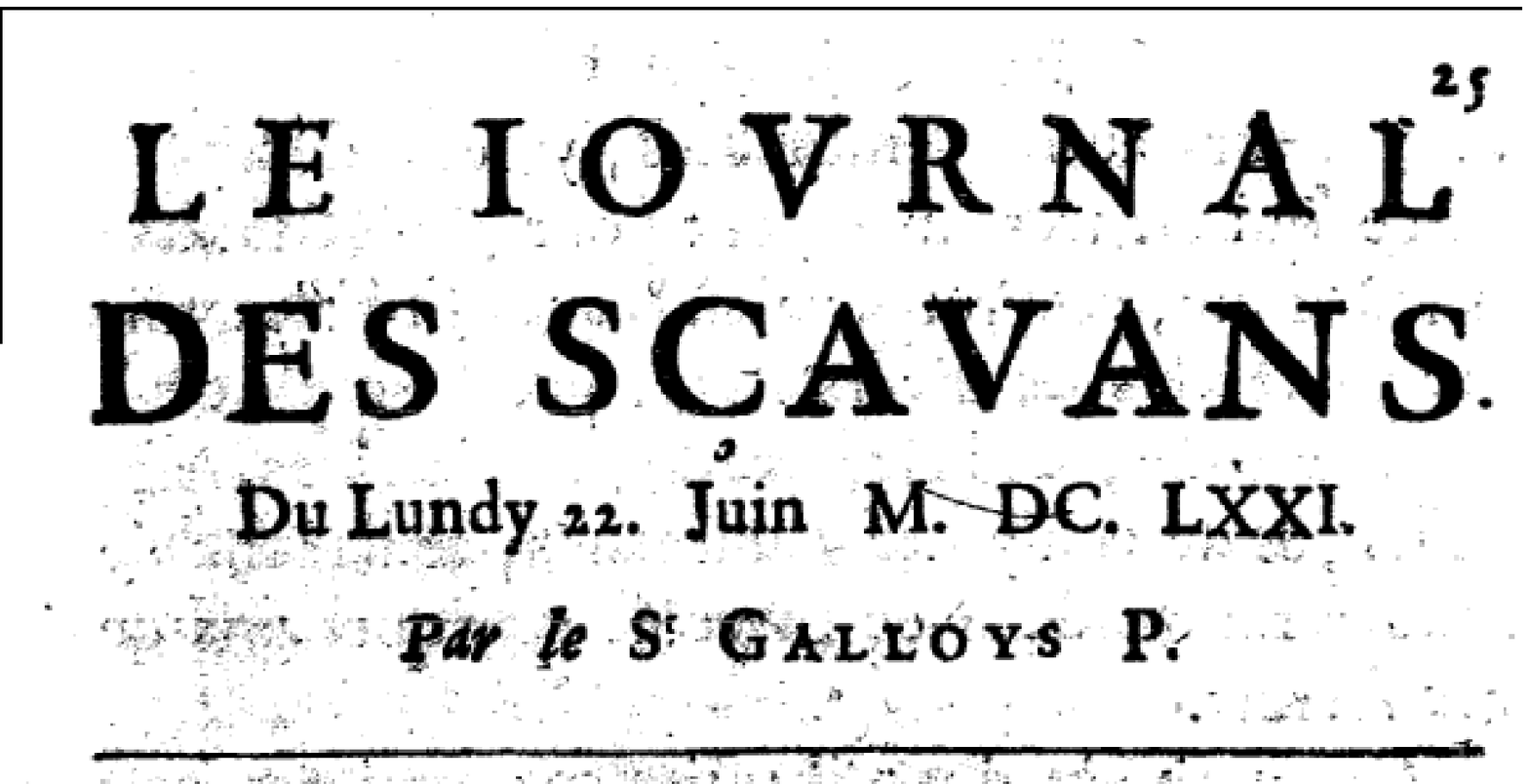}\\
\includegraphics[width=7.8cm, angle=0]{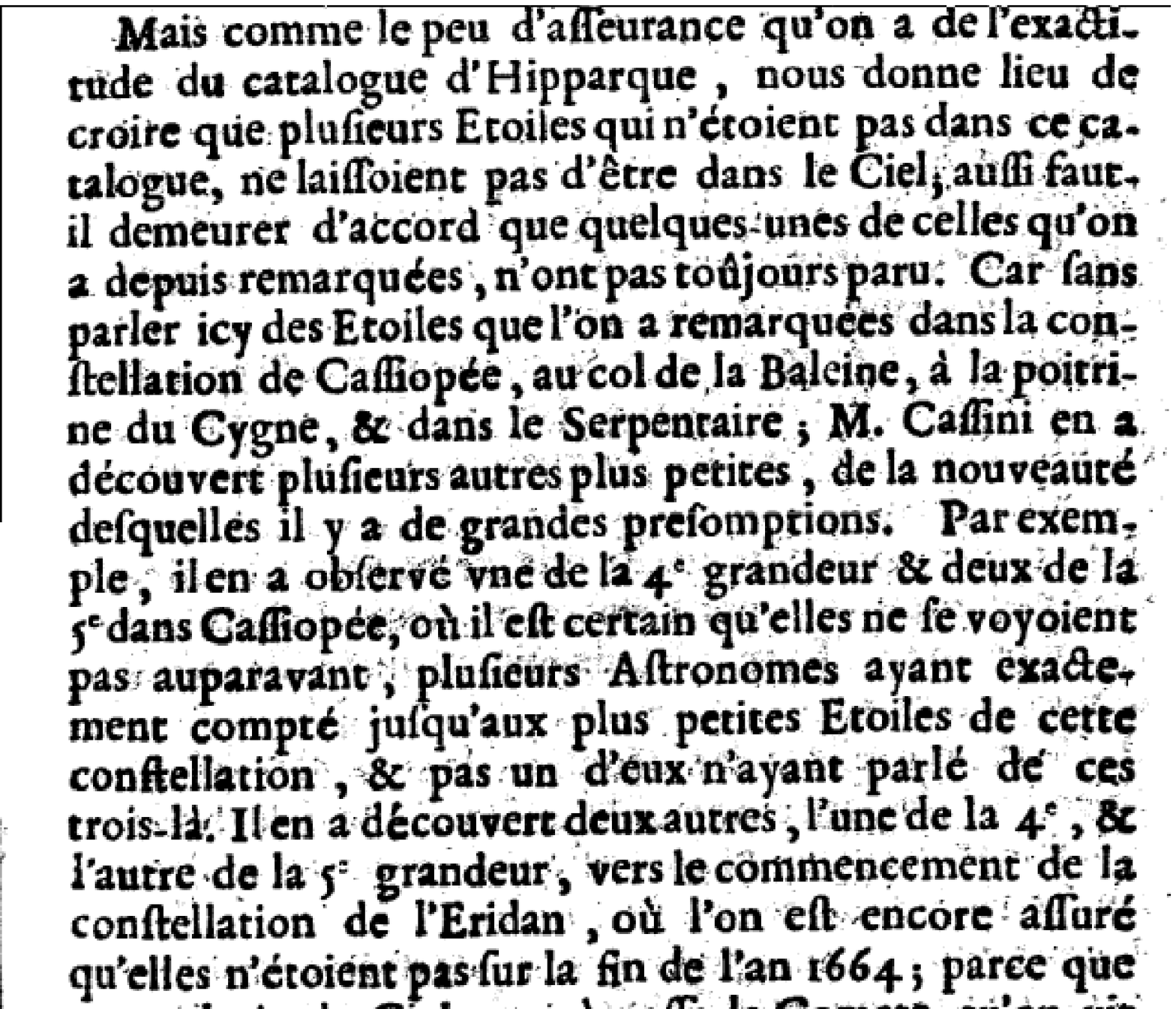}
\caption{Header and excerpts from Jean Gallois's report on new stars, 
published on the Journal des S\c{c}avans in 1671. We argue that 
the 4th-magnitude star in Cassiopeia (never seen before or since) 
could be the Cas A SN.}
\label{figexample}
%\end{center}
\end{figure}

\begin{figure}
%\begin{center}
\includegraphics[width=7.8cm, angle=0]{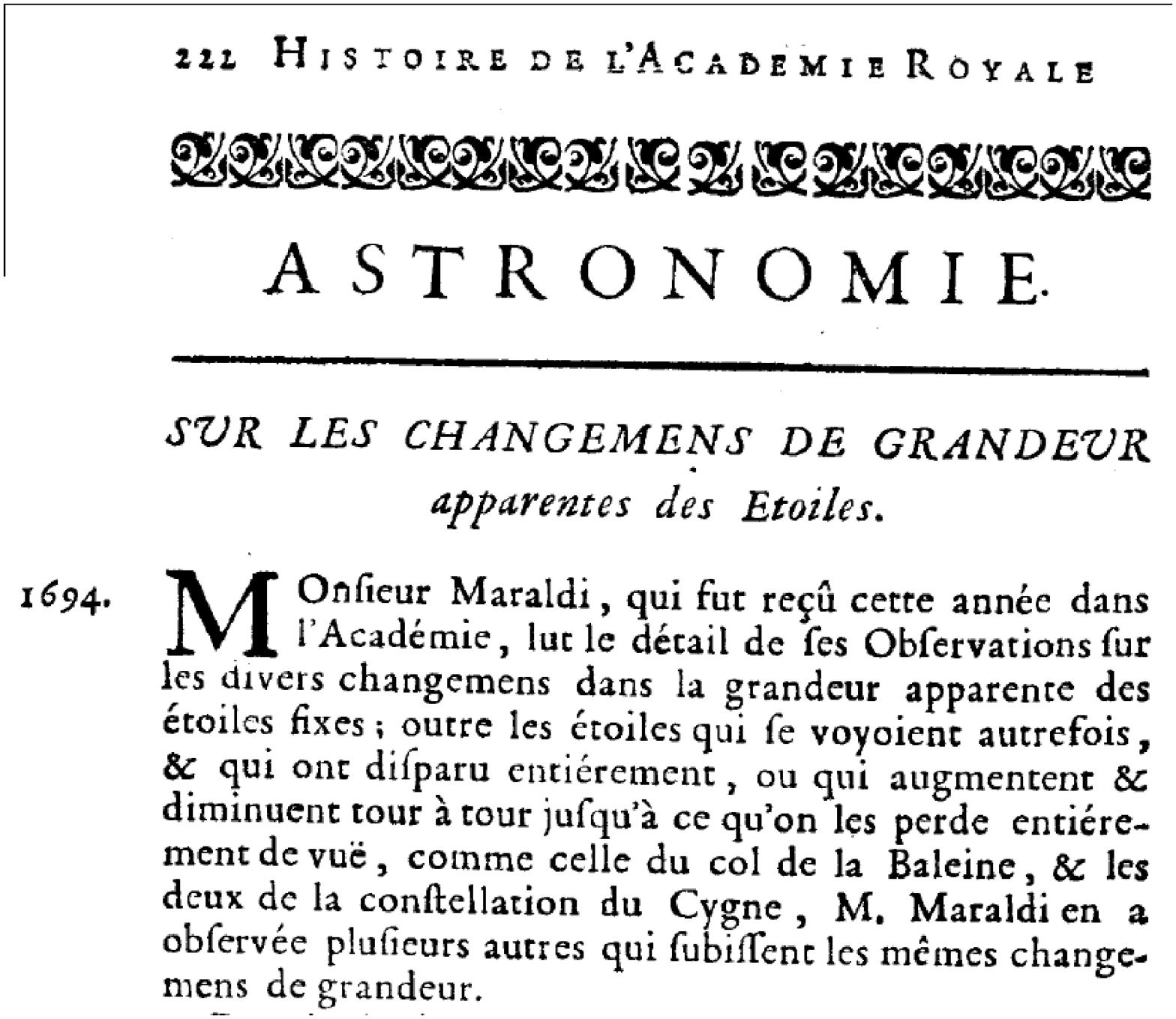}\\
\includegraphics[width=7.8cm, angle=0]{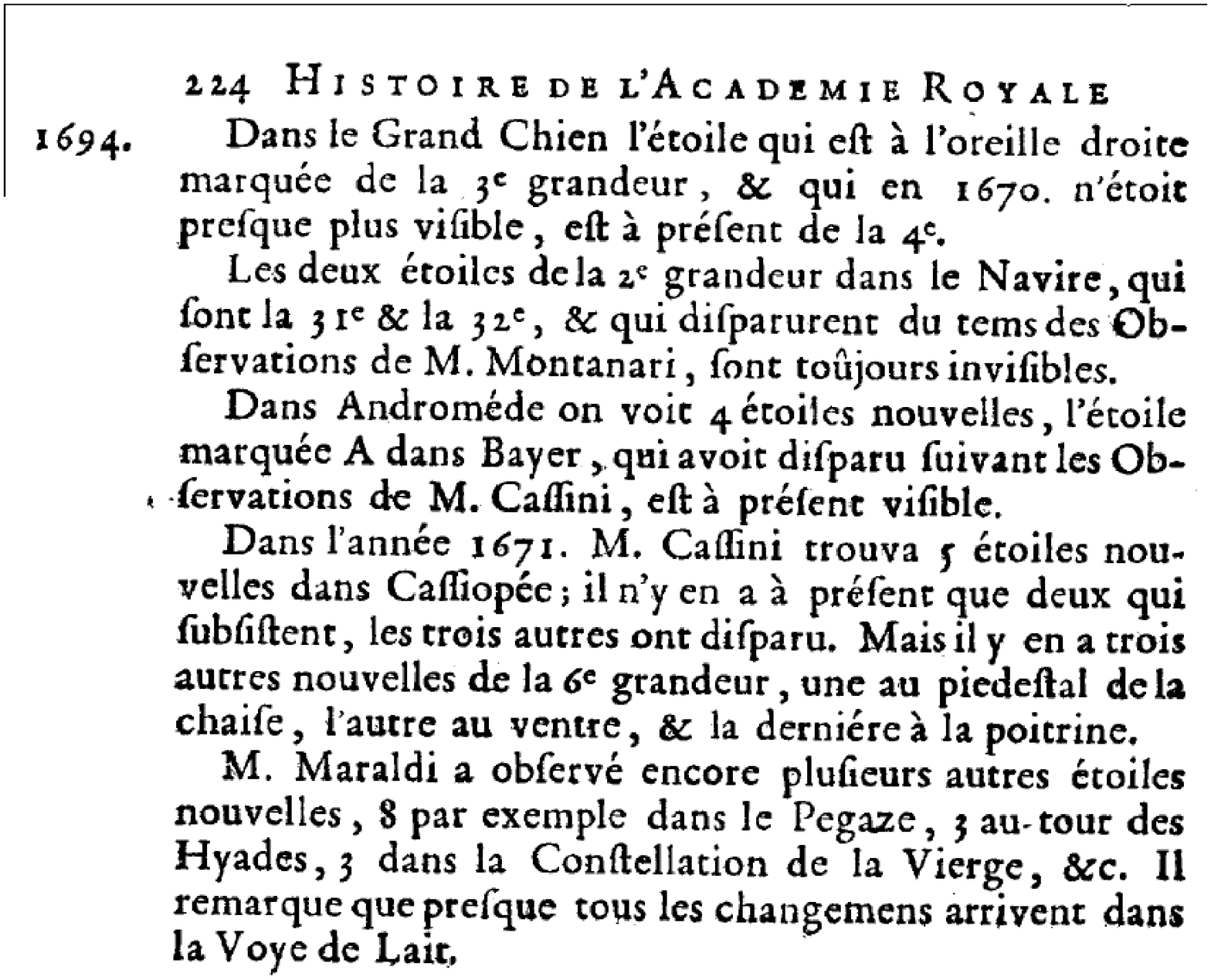}
\caption{Header and excerpts from the Histoire de l'Acad\'emie Royale 
des Sciences, reporting on Maraldi's communication in 1694 that some 
of the new stars found by Cassini in Cassiopeia were no longer visible.}
\label{figexample}
%\end{center}
\end{figure}

\section{The 1671 Cassini report}
Based on the characteristic age extrapolated from the ejecta, as discussed 
in Section 5, we decided to focus our search for further clues between 
about the years 1660 and 1680. By that time, observational astronomy 
had become more advanced in Europe than in East Asia, 
and a 3th--4th mag object would be more likely to be discovered 
and catalogued there. If there was one person in the Old Continent specifically
interested in the study of variable or new stars, that was Gian Domenico 
Cassini (1625--1712).
Cassini and his collaborator and academic successor Geminiano Montanari 
(1633--1687) established the first systematic survey of variable stars 
at the University of Bologna in the late 1660s (after Montanari had 
serendipitously discovered the variability of Algol in 1667). 
Cassini moved to Paris on 1669 April 4, invited by King Louis XIV,  
but continued to pursue this line of research from the newly established 
Royal Observatory in Paris, which opened in 1671.
He was the most renowned and probably most skilled 
astronomer in Continental Europe at the time, enjoying the same prestige 
as the British Astronomer Royal Sir John Flamsteed.

Only a minimal part of Cassini's manuscript 
observation records have been catalogued 
and published; other information on his work is available from second-hand 
reports. We searched and found an article written by the natural 
philosopher and mathematician Jean Gallois (1632--1707), 
on the ``Journal des S\c{c}avans'' 
(a prestigious French literary and scientific journal founded and edited 
by Gallois himself), dated Monday 22 June 1671 \citep{gallois71}.
The topic of the article is variable stars. Gallois writes (page 35): 
"M. Cassini en a d\'ecouvert plusieurs autres [\'etoiles] 
plus petites, de la nouveaut\'e
desquelles il y a de grandes presomptions. Par exemple, il en a observ\'e une
de la $4^e$ grandeur \& deux de la $5^e$ dans Cassiop\'ee, o\`u il est certain
qu'elles ne se voyoient pas auparavant, plusieurs Astronomes ayant
exactement compt\'e jusqu'aux plus petites Etoiles de cette constellation, \&
pas un d'eux n'ayant parl\'e d\'e ces trois-l\`a".
This article is extremely important, because Gallois is very likely reporting 
what he heard directly from Cassini; the two scientists were close friends 
and Gallois used to proof-read the French language of Cassini's scientific 
communications \citep{cassini10}. In summary, Cassini reported 
the discovery of a 4th-magnitude star in Cassiopeia in 1671 
(together with two fainter ones), a star 
that was not accounted for in any previous sky chart. 

This is interesting because we know that all 4th-magnitude stars in Cassiopeia 
had already been discovered and published in sky charts decades before 
Cassini's observations. For example, we examined Johann Bayer's 
Uranometria \citep{bayer03} and compared it with today's Hipparcos 
catalogue \citep{vanleeuwen07}. Bayer's chart is complete down to 
$V = 4.83$ mag. This completeness limit takes into account a couple 
of binary stars that could not be visually resolved at the beginning 
of the 17th century. Notice that some of Bayer's stars (in particular, 
50 Cas, $V = 3.95$ mag) are drawn with the correct location 
and magnitude in his charts but did not receive 
a Greek-letter classification: that is routinely the case for stars 
located outside the ``classical'' boundary of a constellation defined 
by the drawing of its traditional mythological figure. 
We also checked that Bayer's scale of visual magnitudes 
corresponds to today's definition; we found that stars classified as 
``4th magnitude'' correspond to $3.5 \la V \la 4.5$ mag 
in the Hipparcos catalog. Therefore, there should be no ambiguity 
when Cassini mentions the discovery of a 4th-magnitude star.

\begin{figure*}
\begin{center}
\includegraphics[scale=0.5, angle=0]{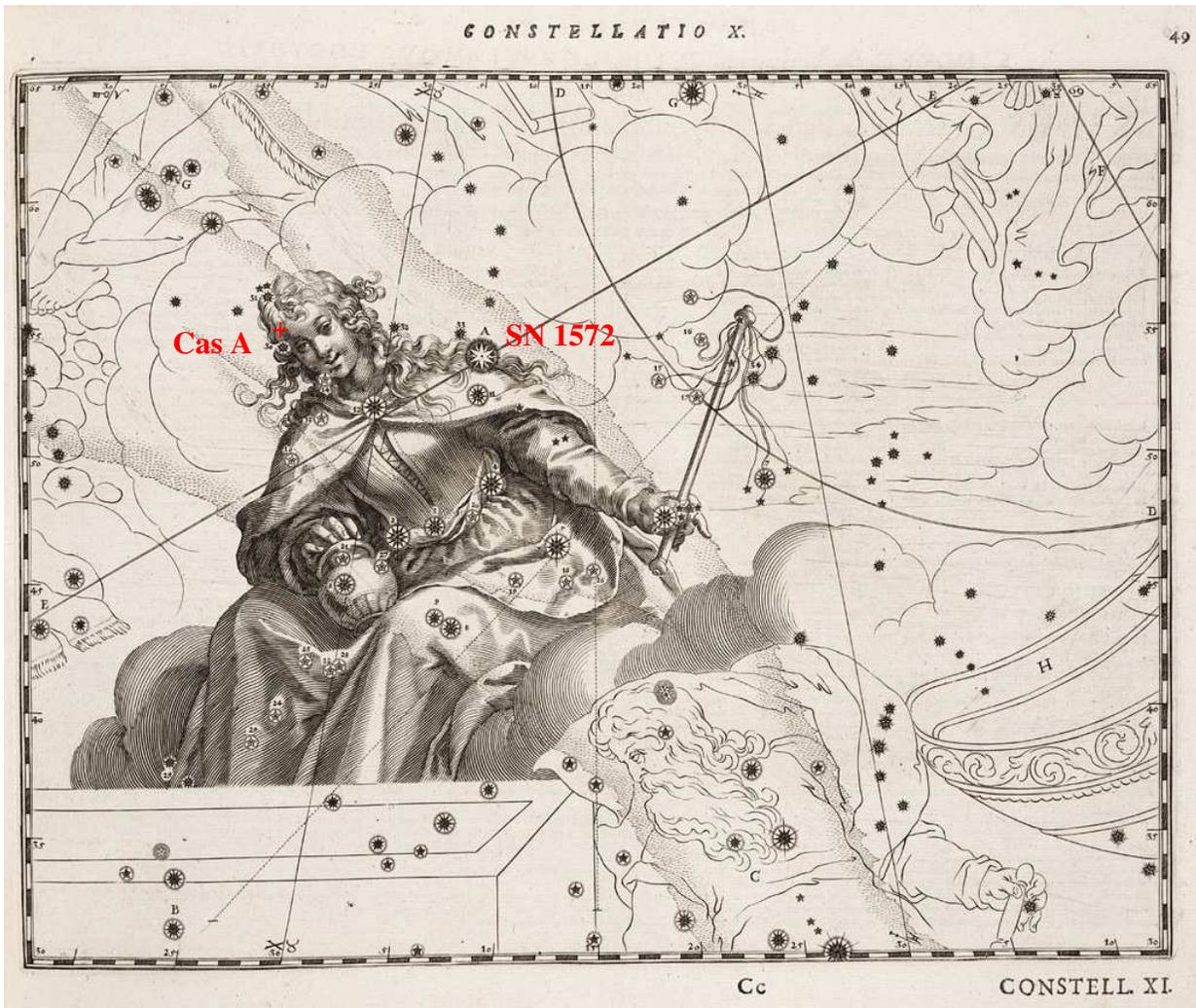}
\caption{An ancient star map: ``St Mary Magdalen alias Cassiopeia'' 
in Schiller's Coelum Stellatum Christianum (circa 1627); source: 
the Linda Hall Library of Science, Engineering \& Technology. 
We overplotted the positions of Cas A and of Tycho's SN 1572.}\label{figexample}
\end{center}
\end{figure*}

\begin{figure*}
\begin{center}
\includegraphics[scale=0.38, angle=0]{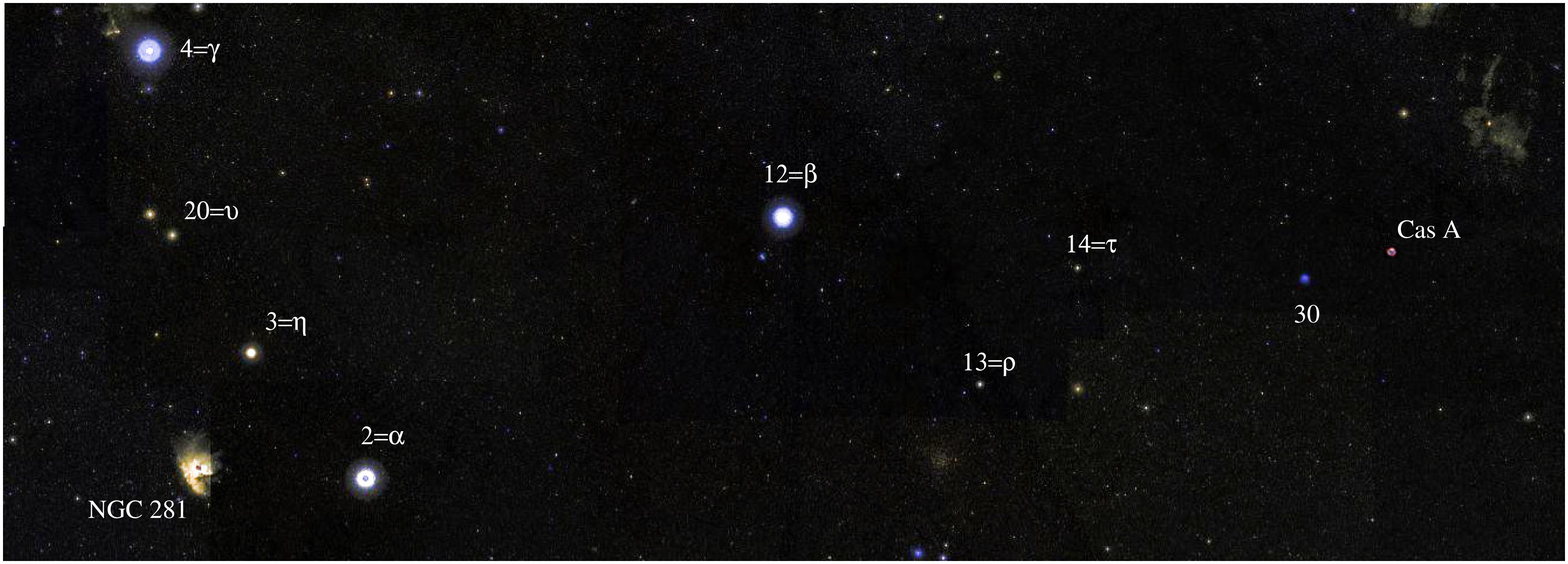}
\caption{A modern sky map, from GoogleSky (circa 2012), 
showing the eastern section of Cassiopeia. 
We labelled some of the stars with Schiller's 
numbers and Bayer's Greek letters. Size of the image: 
$13^{\circ} \times 5^{\circ}.5$. North is up, east to the left. 
Schiller's star Number 30 (present but not labelled 
in Bayer's map) is known today 
as AR Cas, and it may be the object Flamsteed really saw in 1680 
\citep{kamper80,green03}.}
\label{figexample}
\end{center}
\end{figure*}

Twenty-five years after the publication of Uranometria, Julius Schiller 
(1580--1627) worked with Johann Bayer (1572--1625) 
to create a revised and updated stellar atlas 
\citep{schiller27}. 
Schiller did not simply replace all of the pagan constellations with 
Christian figures: he added several new stars discovered after the 
publication of Bayer's charts, and refined their positions.
%For example, this is the first star atlas to include the Great Nebula in Andromeda.
We examined Schiller's map of Cassiopeia and found that it is complete 
down to $V = 4.95$ mag (again, allowing for unresolved double stars).

Cassini was certainly well familiar with both Bayer's and Schiller's charts.
We find it impossible to believe that the best astronomer in Europe 
at the time would not know all the dozen-or-so 4th-magnitude stars 
of Cassiopeia, or that he estimated the wrong brightness 
for his {\it nouvelle \'etoile}. That is even more implausible 
considering that Cassiopeia had probably been the most closely monitored 
region of the sky since the stunning appearance of SN 1572, and that 
Cassini was specifically interested in variable stars and therefore 
in the possibility of a re-appearance of that source. Moreover, 
Cassiopeia is circumpolar from Europe, so it can be observed all year round.
As for the other two new stars observed by Cassini in Cassiopeia in 
or before 1671, 
they were of 5th magnitude, at which level the existing charts were 
not complete; therefore we do not need to invoke special variability for those.

The next reference we found about Cassini's new stars in Cassiopeia 
is from the "Histoire \& M\'emoires de l'Acad\'emie Royale des Sciences''
\citep{maraldi94}. We read there that the Italian-French astronomer 
Giacomo Filippo (Jacques Philippe) Maraldi (1665--1729), inducted into the 
Acad\'emie in 1694 and research staff member at the Paris Observatory since 
1687, presented a report ``On changes of the apparent brightness of stars'' 
("Sur les changemens de grandeur apparentes des Etoiles"). 
Maraldi's communication was read to the Acad\'emie on 1694 December 4.
Secretary Jean-Baptiste Du Hamel (1623--1706) summarized it as such, 
in the Histoire:
"Dans l'ann\'ee 1671 M. Cassini trouva 5 \'etoiles nouvelles dans Cassiop\'ee; 
il n'y en a \`a pr\'esent que deux qui subsistent, les trois autres ont disparu.
Mais il y en a trois autres nouvelles de la $6^e$ grandeur, une au piedestal 
de la chaise, l'autre au ventre, \& la derni\`ere \`a la poitrine".

A further mention of the variable stars in Cassiopeia observed by Cassini 
is found in his son Jacques Cassini (Cassini II)'s Elemens d'astronomie 
\citep{cassini40}, although he appears to summarize the information 
previously reported by Gallois and Maraldi: 
"Outre ces \'{e}toiles dont on vient de faire le rapport, 
mon pere en a d\'{e}couvert plusieurs autres plus petites, qu'on pr\'{e}sume 
\^{e}tre nouvelles. Par exemple, il en a observ\'{e} une de la quatrieme 
grandeur, \& deux de la cinquieme, dans la constellation de Cassiop\'{e}e, 
o\`{u} il est certain qu'elles ne se voyoient pau auparavant, n'y ayant 
aucun Astronome qui en ait fait mention, quoiqu'il y en ait eu plusieurs 
qui ayent exactement compt\'{e} jusqu'aux plus petites \'{e}toiles 
de cette constellation. En 1671, il trouva cinq nouvelles \'{e}tolies 
dans la Cassiop\'{e}e, dont trois avoient disparu".

We do not know the coordinates of the transient 4th-magnitude star 
discovered by Cassini in Cassiopeia, nor of the other variable ones.
Cas A is outside the traditional mythological boundaries of Cassiopeia, 
located between Cassiopeia and Cepheus: this may explain why 
its location was not reported more precisely at the time.
Moreover, we should not assume (based on the words used by Gallois 
and Maraldi) that the first observation of the transient star in Cassiopeia 
occurred precisely in the first half of 1671. Cassini was known to observe 
his targets meticulously for years, sometimes, before reporting a result 
or a new discovery. This caveat is important in view of the nitrate 
spike detected in the 1667 ice layers. On the other hand, 
on 1668 July 2, the Journal des S\c{c}avans reported on the 
``Apparizioni celesti dell'anno 1668 osservate in Bologna da Gio.~Domenico 
Cassini Astronomo dello studio publico'' \citep{gallois68}, 
in which other variable stars are mentioned, but not those in Cassiopeia.

To make further progress, we need to access and search through 
the original records of Cassini's observations, especially those 
between 1669 and 1671.
% si trova al di fuori delle figure tradizionali, tra Cassiopea e Cefeo:
%se e' stata osservata la sua progenitrice, cio' complicava la segnalazione
%della posizione nel modo "letterario" allora usato; e' possibile che cio'
%spieghi la vaghezza delle citazioni di cui sopra.
Such documents are still kept in the archives of the Paris Observatory, 
and have never been catalogued or scanned. As we send this paper to press, 
we are still negotiating the possibility of accessing those archive 
in Paris for further research. Hopefully, we will report on this 
in follow-up work. But for now, we can already propose that Cassini's 1671 
reported observation of a 4th-magnitude transient star in Cassiopeia 
is at least as strong a candidate for Cas A as Flamsteed's 1680 observation. 

%But what if they were both right? What if there was a precursor event in 1671 and then a rebrightening of the same object in 1680? Until a few years ago, the idea of 

\section{Conclusions}
We examined recent claims that the Cas A SN might be identified 
with a 1630 ``noon-star'' reported in the English literature. We strongly 
disfavour this possibility, based on the expected brightness of a Type-IIb SN 
(too faint to be seen in daylight), the extrapolated motion of the ejecta 
(inconsistent with any explosion date earlier than at least 1650), 
the lack of any scholarly reference to the event. We detail strong evidence 
that there was a bright comet in 1630 June, but we found no evidence 
to determine whether it could be visible in daylight.
We also found no record of day-time guest stars or broom stars 
consistent with a 1630 Cas A SN in the Japanese archives.
Based on the motion of the ejecta, we focused our search 
for the Cas A progenitor to the years between about 1660 and 1680, 
and to the astronomers 
who would have been most likely to notice it: Gian Domenico Cassini 
and his collaborators, who were doing pioneering work on variable stars.
We found a report about a 4th-magnitude star (that is, with the brightness 
expected for the Cas A SN) discovered by Cassini in Cassiopeia 
in or shortly before 1671 (the same epoch inferred for the event by 
\citealt{thorstensen01}), which was not seen before or since. 
We argue that this source could be the long-sought SN, 
but further research in the 
original observing logs (kept at the Paris Observatory) 
is needed to determine the discovery date and coordinates 
of the transient Cassiopeia object observed by Cassini.

%Arcavi, I. 2011ApJ...742L..18A   2011dh
%A. Pastorello 2008, MNRAS, 389, 955   2008ax
%Bembrick, C.; Pearce, A.; Evans, R.2002IAUC.7804....2B   2001ig
%Qiu, Y., Li, W., Qiao, Q., & Hu, J. 1999, AJ, 117, 736  SN 1996cb
%Reed, J. E., Hester, J. J., Fabian, A. C., & Winkler, P. F. 1995, ApJ, 440, 706
%Schmidt, Brian P. 1993Natur.364..600S   1993J

\section*{Acknowledgments} %If needed
We thank Fabrizio B\`onoli, Valeria Zanini, Suzanne D\'{e}barbat, 
Nicola Masetti, Naomi Mockford and an anonymous referee 
for their suggestions and comments. 
We thank Toru Hoya and Hidenori Ouchi for their invaluable help 
with the Japanese records at the Historiographical Institute of the
University of Tokyo.

%\end{multicols}

\end{document}